\documentclass{article}
\usepackage{epsfig}
\def\slash#1{\setbox0=\hbox{$#1$}#1\hskip-\wd0\dimen0=5pt\advance
      \dimen0 by-\ht0\advance\dimen0 by\dp0\lower0.5\dimen0\hbox
         to\wd0{\hss\sl/\/\hss}}
\begin{document}
\begin{center}
{{\bf \Large Puzzles of excited charm meson masses}}

\bigskip 

B. Ananthanarayan$^a$\\
Sunanda Banerjee$^b$\\
K. Shivaraj$^a$\\
A. Upadhyay$^a$

\bigskip

$^a$Centre for High Energy Physics, Indian Institute of Science,\\
Bangalore 560 012

\medskip

$^b$Department of High Energy Physics, Tata Institute of Fundamental Research,\\
Homi Bhabha Road, Mumbai 400 005.

\end{center}
\bigskip

\begin{abstract}
We attempt a comprehensive analysis of the low lying charm meson
states which present several puzzles, including the poor
determination of  masses of several non-strange excited
mesons.  We use the well-determined masses of the
ground states and the strange first excited states to   
`predict' the mass of the non-strange first excited state
in the framework of heavy hadron chiral perturbation theory,
an approach that is complementary to the well-known analysis of
Mehen and Springer.  This approach points to values for
the masses of these states that are smaller than the 
experimental determinations.  We provide a 
critical assessment of these mass measurements and point out the 
need for new experimental information.
\end{abstract}

\section{Introduction}
The Particle Data Group in its latest Review of Particle Properties~\cite{PDG}
lists several low-lying charm meson states, the ground state being
the $J^P=0^-,\, 1^-$ states, and the first set of excited states
corresponding to $J^P=0^+,\, 1^+$.  The latter present some
puzzles: the strange mesons known as $D_s(2317)$
and $D_s(2460)$ are said to be lighter than expected from
constitutent quark model predictions (for reviews, see 
e.g.~\cite{CFF,vijande}).
The observation of $D_s(2317)$ was first reported in~\cite{2317}
and the confirmation reported in~\cite{2460}, the latter also
reporting the discovery of $D_s(2460)$.  The non-strange counterparts
of these states had been reported earlier in 
refs.~\cite{hep-ex/9908009,hep-ex/0307021} which are
denoted $D^0_0(2308)$ and $D^{0'}_1(2438)$ (the last number 
resulting from an average over BELLE~\cite{hep-ex/0307021} and CLEO~\cite{hep-ex/9908009}numbers).
The observation of charged non-strange mesons with $0^+$
quantum numbers were reported by the FOCUS collaboration~\cite{PHLTA.B586.11}
at a mass of about 2460 MeV.  
In view of a conflict with the measurements of the strange meson 
masses, it has been disregarded in some theoretical
treatments.  No observation
of the corresponding charged non-strange meson with $1^+$ has
been reported so far.  A summary is provided in the diagram
given in Fig. 1.

\begin{figure}[ht]
\includegraphics[width=12cm]{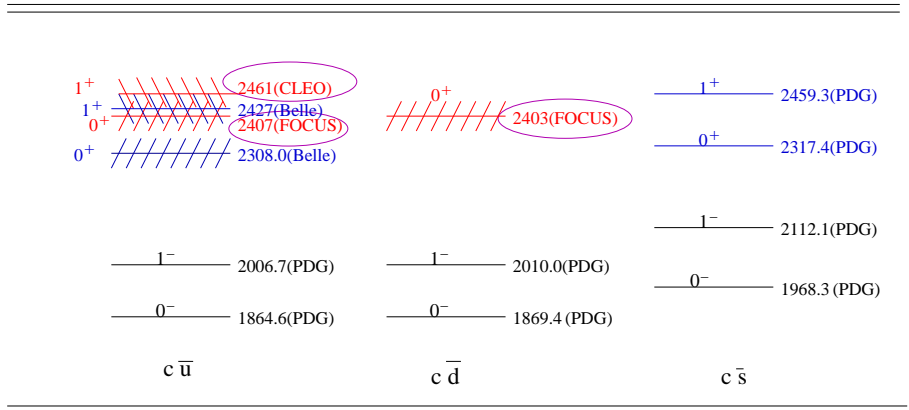}
\caption{Representation of the masses in MeV of
the charm mesons with their quantum number assignments,
and the names of the experiments for determinations
associated with the non-strange excited states.}
\end{figure}

The masses have been analyzed in great detail in the framework of
heavy hadron chiral perturbation theory              
by Mehen and Springer~\cite{MSmasses}, including
one-loop chiral and $O(m_c^{-1})$ corrections. 
This is a hybrid framework that exploits the
chiral symmetry associated with the lightness of the
u-, d- and s-quarks and the heavy quark symmetry that
is realized in the limit of the mass of the heavier quark, {\it viz.}
b or c, becoming very large.  Of the conclusions drawn
in ref.~\cite{MSmasses}, a notable one is that due to the scarcity
of experimental information, it was hard to fix either the tree-level
parameters or the the coupling constants entering the one-loop corrections
to any significant accuracy.  

More recently, best fit values for
three of the coupling constants, $g,\, g', h$
have been reported, taking into account chiral corrections
to the strong decays of these positive and negative
parity states~\cite{hep-ph/0606278}.  These determinations are 
in agreement with the orders of magnitudes              
considered earlier in the literature~\cite{hep-ex/0108043,hep-ph/9706271}.
In particular, note however, $g'$ continues to be a parameter
for which there is no direct experimental evidence and
results only from a combined fit to the information on
decays and other sources of information.  While carrying 
out our numerical fits, we will confine these parameters
to lie in the range compatible with the numbers presented
in ref.~\cite{hep-ph/0606278}.  

Our main motivation is to find the masses of the non-strange
excited states from the observed experimental values of all the
ground states and excited strange mesons only.  
In the present work, we employ the expressions presented
in the comprehensive work of Mehen and Springer~\cite{MSmasses}.  
Our approach is in a sense complementary to that of
ref.~\cite{MSmasses}, where the masses of the excited 
non-strange states
are also used in their fits.  One of the main
differences in our work  stems not only from
constraining the range of $g$ and $h$ to be 0 to 1, but also
requiring other parameters to lie in a narrow range satisfying 
experimental mass determinations. Note however, that we will consider only 
those fits as proper in which $g'$ is significantly smaller
in magnitude in comparison with $g,\, h$.  This is in accordance
with the observations presented in ref.~\cite{BFP}, which are
based on arguments coming from QCD sum rules and relativistic
as well as non-relativistic quark model. 
Another ingredient in 
which we differ from their work is that the mass of the s-quark is chosen to be 
130 MeV, which differs from the value 90 MeV used in 
ref.~\cite{MSmasses}.  The difference stems from the renormalization
group evolution of the $\overline{\rm MS}$ mass from the scale
2 GeV down to the present renormalization scale of 1 GeV.
In practice, it is found that the masses are not very sensitive to 
this choice.
                                      
We find that the corresponding predictions for the non-strange
states can be in conflict with experiment. Our fits are able to produce
lower lying positive parity states with masses less than experimental 
observations. If we consider that masses of strange D-mesons are well 
determined than this would be the prediction of ref.~\cite{BFP}. In 
ref.~\cite{BFP} it is
shown that the difference $(m_{D_0^*}-m_D)-(m_{D_{s0}^*}-m_{D_s})$ 
has to be less than zero, after the inclusion of chiral
corrections. Here we explicitly calculate the masses
of non-strange mesons, and show that they indeed satisfy the above relation. 
We must caution, however, that since there are a larger number of
parameters than there are observables, fitting the masses of the
(poorly determined) non-strange masses can also yield a perfectly
natural set of parameters for the theory.

We then turn to a study of all the experiments that have reported
the masses of the non-strange excited states.  We find that the determination
of the masses of the  neutral non-strange
mesons do not appear to be consistent.  In addition, in the 
charged non-strange sector the
experimental situation remains grossly unsatisfactory, as only the
FOCUS collaboration has reported the observation of the $0^+$ state
with fairly large errors and a central value higher than the
corresponding strange meson, in conflict with SU(3) predictions, 
while no experiment has reported the observation of the $1^+$ state
in this sector.  A resolution of the situation could emerge if an
independent experiment, e.g., BABAR could observe all these states
and carry out a measurement of the masses of these mesons.
Also the CLEO-c experiment could attempt to observe all the states
discussed here at high precision.

\section{Results from heavy hadron chiral perturbation theory}

In the framework of heavy hadron chiral perturbation theory
chiral corrections and corrections due to the finite mass of
the charm quark are also accounted for.  
Our analysis uses the expressions for masses of the charm mesons
presented in ref.~\cite{MSmasses}.  The masses are expressed in
terms of a formula which reads, for the residual masses,
\begin{equation}
m^0_{R_a}=\delta_R + {n_J\over 4} (\Delta_R + \Delta^{\sigma}_R \overline{m}+
				   \Delta^{(a)}_R m_a)+\sigma_R \overline{m}
				+ a_R m_a + {g_R^2\over f^2} c^{R_a} K_1+
					    {h^2\over f^2} c^{R_a} K_2,
\end{equation} 
where $R$ is an index that labels the ground state ($H$) and 
excited state ($S$),
each of the ground and excited states having members corresponding to
$J=0,\, 1$, with $n_0=-3,\, n_1=1$, the index $a$ labels the light
flavour and runs over $u,\, d,\, s$, the functions $K_1$ and $K_2$ 
are the chiral loop functions, and the $c^{R_a}$ are coefficients
listed in \cite{MSmasses} and $g_H=g,\, g_S=g'$ in the notation
therein.  The relevant Lagrangian is reproduced 
for completeness in the Appendix.   

Some insight can be obtained on the coefficients of interest in
certain limits.  For instance, in the heavy quark effective field
theory limit, one can make the identification $\delta_R\to 
\overline{\Lambda}_R + \lambda_1^R/(2 m_Q)$, 
$\Delta_R \to \lambda_2^R/(2 m_Q)$.  The heavy quark effective
theory constants on the right-hand side of each of these relations
have been estimated for the ground-state, namely for $R=H$.
The reaction $b\to s \gamma$ measured by the CLEO
collaboration~\cite{hep-ex/0108032}
requires the constant $\overline{\Lambda}_H \simeq 0.35 {\rm GeV}$.
Fits to b-decays yield 
$\lambda_1^H\simeq -0.20 {\rm GeV}^2$~\cite{bdecays},
while measurements on the lattice yield for 
$\lambda_2^H\simeq 0.10 {\rm GeV}^2$~\cite{hep-lat/0305024}.
Note that in the treatment of ref.~\cite{BFP}, the hyperfine splitting
of the odd-parity ground state, and that of the even-parity
first excited state has to be equal.  The lifting
of the degeneracy of the hyperfine splitting is
considered to be due a term arising from the next to leading order
in the heavy quark effective theory.  In the treatment of Mehen 
and Springer, the leading and next to leading order effects are
both taken into account, both through the lumping of the effects
into the $\delta_H$, as well as in the hyperfine splitting effects.
The hyperfine splitting itself is of interest, since it seems
to be equal in the ground as well as the first excited states.
This is the subject of a recent study, ref.~\cite{hep-ph/0701245}.
We now describe the results from our fits in the foregoing.

\subsection{Results from constrained fits}
In our numerical work we use the values for the quark masses of
$m_u=m_d=4,\, m_s=130$ MeV.  The latter mass differes significantly
from that used in ref.~\cite{MSmasses}; the change coming primarily
from our having to use the mass of the s-quark at 1 GeV, related to
the mass given as 95 MeV at 2 GeV, these being related by a factor
of $\simeq 1.35$.

In our fit, we use six experimentally determined masses,
{\it viz.} four from the strange
sector (both negative and positive parity states) and two from
the non-strange sector (negative parity states), in the iso-spin limit. 
The loop corrections
depend upon 11 parameters: $g,\, g',\, h, \, \delta_H,\, \delta_S$,
$\Delta_H$, $\Delta_S$, $a_H$, $a_S$, $\Delta_H^{(a)}$, $\Delta_S^{(a)}$. 
Since there is a surplus of parameters, a unique fit is not possible. 
Nevertheless, we could
find good fits by constraining the parameters in well-motivated ranges.
Special mention may be made to the values presented in
ref.~\cite{hep-ph/0606278,hep-ph/0701245}.  Note that in 
ref.~\cite{hep-ph/0701245}, heavy hadron chiral perturbation theory
with only the ground states has been considered.

As an illustration we present the results from one of our fits
by requiring that the parameters $g,\, h$ lie in the range
(0,1) and $g'$ in the range (-0.5,0.5).  This yields, in the
notation of ref.~\cite{MSmasses}, 

\begin{center}
\begin{tabular}{ll}\label{fit1}
$\delta_H = 314.7${\rm MeV} &  $\delta_S = 688.6 $ {\rm MeV} \\
$\Delta_H = 149.9 $ {\rm MeV} & $\Delta_S = 725.1 $ {\rm MeV} \\
$a_H = -5.071$ & $a_S=-6.03$ \\
$\Delta^{(a)}_H = -8.883 $ & $\Delta^{(a)}_S = -9.544 $
\end{tabular}
\end{center}
with $g,\, g', h$ being 0.905, 0.001, 0.998 respectively. From this fit, we 
obtain for the non-strange $0^+$ state the mass of 2155.7 MeV and for $1^+$ 
2395.1 MeV.

From the experimental  values of non-strange ground states and strange ground 
as well as excited states, we can observe that $\Delta_H $ and $\Delta_S $ are 
of the order of 140 ${\rm MeV}$.  Therefore, we allow each of them to vary
from 100 to 200 MeV.  From the difference between different 
parity states in the strange sector we can see that $\delta_H - \delta_S 
\leq$ 
400 ${\rm MeV}$, and therefore each of them is allowed to vary
between 100 and 500 MeV. Moreover the mass difference between strange 
and non-strange ground states, which is of the order of 100 ${\rm MeV}$, gives 
$a_H(m_s - m_{u/d})\sim 100$ leading to $a_H$ to be order of unity. 
Similarly
we consider all other chiral contribution to be of the order of unity. 
Consequently, we allow these parameters to vary from -2 to 2.
The values of $g,\, g', \, h$ are also chosen to be of the orders of
magnitude given in ref.~\cite{hep-ph/0606278}. 
With 
the above constraints to the parameters we get several fits and a few of them 
are given below. 

\begin{center}
\begin{itemize}
\item
$\delta_H = 169.2 \pm 0.5 $ {\rm MeV} \hspace*{0.8cm}  $\delta_S = 345.4 \pm 0.7 $ {\rm MeV} \\
$\Delta_H = 200.3 \pm 0.5 $ {\rm MeV} \hspace*{0.8cm} $\Delta_S = 120.4 \pm 1.1 $ {\rm MeV} \\
$a_H = 1.522 \pm 0.005 $ \hspace*{1.2cm} $a_S=0.508 \pm  0.018 $ \\
$\Delta^{(a)}_H = -1.231 \pm 0.005 $ \hspace*{0.8cm} $\Delta^{(a)}_S = 0.193 \pm 0.015 $ \\
$|g| = 0.66 \pm  0.01 $ \hspace*{0.5cm} $|g'| = 0.03 \pm 0.01 $ \hspace*{0.5cm} $|h| = 0.42 \pm 0.01 $

\item 
$\delta_H = 184.9 \pm 3.6 $ {\rm MeV} \hspace*{0.8cm} $\delta_S = 368.1 \pm 2.4 $ {\rm MeV} \\
$\Delta_H = 196.5 \pm 3.2 $ {\rm MeV} \hspace*{0.8cm} $\Delta_S = 119.4 \pm 1.4 $ {\rm MeV} \\
$a_H = 1.534 \pm 0.006 $ \hspace*{1.2cm} $a_S=0.453 \pm 0.039 $ \\
$\Delta^{(a)}_H = -1.242 \pm 0.027 $ \hspace*{0.8cm} $\Delta^{(a)}_S = 0.189 \pm 0.022  $ \\
$|g| = 0.68 \pm 0.01 $ \hspace*{0.5cm}  $|g'| = 0.01 \pm 0.04 $ \hspace*{0.5cm} $|h| = 0.32 \pm 0.02 $ 

\item
$\delta_H = 159.2 \pm 3.7 $ {\rm MeV} \hspace*{0.8cm} $\delta_S = 350.6 \pm 2.7 $ {\rm MeV} \\
$\Delta_H = 194.4 \pm 3.2 $ {\rm MeV} \hspace*{0.8cm} $\Delta_S = 145.9 \pm 1.4 $ {\rm MeV} \\
$a_H = 1.524 \pm 0.008 $ \hspace*{1.2cm} $a_S=0.423 \pm 0.043 $ \\
$\Delta^{(a)}_H = -1.148 \pm 0.028 $ \hspace*{0.8cm} $\Delta^{(a)}_S = -0.01 \pm 0.02 $ \\
$|g| = 0.65 \pm 0.01 $ \hspace*{0.5cm} $|g'| = 0.05 \pm 0.03 $ \hspace*{0.5cm} $|h| = 0.45 \pm 0.02$ 
\end{itemize}
\end{center}

  All these fits give mass values for $0^+$ state to be 
in the range 2200-2250 MeV,
and $1^+$ to be in the range 2335-2375 MeV. Even though there are yet other 
regions of parameter space in which these mass values can change, 
here it is our main aim to point out that these values do not contradict 
any present day theory.  In such regions of parameter space, the
independent parameters of the theory would  be in very
different ranges compared to those considered here.  

\section{The Experiments}
We present here a discussion on the experiments that have
reported the excited even-parity states.  The first reports
for the non-strange states come from CLEO~\cite{hep-ex/9908009}
which saw the $l=1$, $1^+$ state.  The central value reported
here is 2461 MeV, which renders state heavier than the corresponding
strange state.  The situation was mitigated by the measurement
of the mass of this state with a central value of 2427 MeV by
the BELLE collaboration~\cite{hep-ex/0307021}.  The average
of the two experiments has been used to label the state
as $D^0_0(2438)$~\cite{CFF}.  The latest Review of Particle 
Properties~\cite{PDG}  
does not use the CLEO data, and instead labels the state as
$D_1(2420)$.  The corresponding $l=0$ state which does not
conflict the predictions of $SU(3)$ symmetry is
seen only by the BELLE collaboration which reports a central
value for the mass 2308 MeV~\cite{hep-ex/0307021}.
On the other hand, the FOCUS collaboration has reported
signals for both the charged as well as neutral non-strange
$0^+$ states, a little above 2400 MeV, and are thus heavier
than their strange counterparts.  These measurements, therefore,
have been rejected on theoretical grounds.
[Despite this, the latest Review of Particle Properties
labels this state $D_0^*(2400)^0$ with a mass of 2352 MeV
averaged over the BELLE and FOCUS masses, albeit with an
uncertainty of 50 MeV.]
The approach adopted in the work here points to a conclusion that
that the masses 2308 and 2420 are still somewhat large
to be the non-strange counterparts of the strange states at 2317 and 2460.  
Nevertheless, since the theory has more parameters than observables,
successful fits can still be found with these masses, ref.~\cite{MSmasses}.

The FOCUS experiment is based on the interaction of high energy
photons with a fixed target.  It has reported mass and width
measurements of both the neutral as well as the charged non-strange
$0^+$ states, the masses of both begin nearly degenerate as required
by iso-spin invariance, but at a mass higher than the strange
counterpart.  This paradoxical situation has not been rectified
and theorists have simply rejected these measurements without comment.
It would desirable for the collaboration to carry out a reanalysis
of their data to assist in sorting out the difficulties of the
charm meson mass spectra discussed here.

The CLEO experiment has determined the mass of the neutral non-strange
$1^+$.  The data is gathered from $e^+ e^-$ collisions and yields
a mass that is not been included in the latest PDG.  To our knowledge
there is no further data from the collaboration in the past six years
which clarifies the situation, since their measurement is significantly
higher than the central value from BELLE, and is slightly greater than the
mass of the strange counterpart.

Thus, we are left only with the measurement of the neutral non-strange
$0^+$ and $1^+$ masses from the BELLE collaboration.  This data
gathered from the asymmetric B-factory is the only set that does not
openly come into conflict with SU(3) predictions.  Despite this,
the mass determinations of these indicate that they cannot be 
comfortably accomodated into the existing theories as the expected splitting 
from the strange
counterparts would be of the order of 100 MeV.  Our finding with
the constrained fits  also suggest that this can be the
order of the splittings, while the BELLE measurements show that
the splitting is much less.

To summarize, the experimental determinations of the non-strange excited
neutral mesons remain too high and are not necessarily self-consistent,
although all the predicted states have been observed.  In the corresponding
charged sector, only the $0^+$ has been seen, but with a mass in
contradiction with SU(3) prediction, while the $1^+$ has not
been seen at all.
It is our view that an independent experiment, such as
the BABAR collaboration could contribute to
the resolution of the discrepancies here by searching for the states of 
interest.

\section{Discussion and Conclusions}
In this work we have revisited the issue of the masses of lowest-lying
even-parity excited states in the open charm system.  We have worked
in the framework of heavy hadron chiral perturbation theory, including
the $O(m_c^{-1})$ corrections, namely the framework employed in
ref.~\cite{MSmasses}.  The key differences is that we now constrain
the values of the parameters $g,\, g', h$ to the values that are
determined from the decays.  Furthermore we have imposed the requirement
that $m_s\simeq 130$ MeV, and have required the parameters determining
the tree-level hyperfine structure to be in a range determined by
the well-established states.  We find that the masses for the non-strange 
states that we determine can be lower than
the numbers obtained by the BELLE collaboration.  We conclude by noting
that a corroboration of these numbers from other experiments is
imperative.  It must be added that as the number of parameters
exceeds the number of observables, it would be possible to find
natural fits to the parameters even with the present numbers,
as in ref.~\cite{MSmasses}.  
Also of interest is the possibility of understanding these masses
in Regge framework, see ref.~\cite{hep-ph/0703278}.
Finally,
an experimental determination of the charged non-strange even parity
states would significantly clarify the situation.  A determination
of the constants $\lambda_1^S$ and $\lambda^S_2$ on the lattice would
also contribute significantly to the understanding of the 
spectrum\footnote{We thank R. R. Horgan for a discussion on this
subject.}.

\bigskip

\noindent
{\bf Acknowledgements}: We thank the Department of Science and Technology,
Government of India,
for support.  AU is supported under the Project No. 
SR/S2/HEP-12/2004/8.8.05.  We thank the
Council of Scientific and Industrial Research for support.  We thank
R. Springer and T. Mehen for correspondence.  We also thank
J. Goity, R. Patel, R. R. Horgan for discussions.  Finally, BA
thanks the organizers of the workshop ``From Strings to LHC" held
in Goa, January 2-10, 2007 for their hospitality,
when part of this work was done, with special
thanks due to K. Sridhar for providing a convivial atmosphere.

\bigskip

\noindent{\bf Appendix:  Effective Lagrangian for Heavy Hadron Chiral
Perturbation Theory}

In the heavy quark limit, spin of the heavy quark decouples from that of light 
degrees of freedom. Hence the pseudoscalars and vector mesons in the ground
state of D-meson system become degenerate. In this case it is 
convenient to introduce a single field for the ground state doublet as
\begin{eqnarray*}
& \displaystyle
H_a = {1 + \slash{v} \over 2} (H^{\mu}_a \gamma_{\mu} - H_a \gamma_5) \, , &
\end{eqnarray*}
Similarly for the first excited douplet
\begin{eqnarray*}
S_a = {1 + \slash{v} \over 2} (S^{\mu}_a \gamma_{\mu} \gamma_5 - S_a )\, ,
\end{eqnarray*}

Now the Lagrangian that describes the dynamics of mesons with heavy-light  
combination is given by

\begin{eqnarray*}\label{lagrang}
& \displaystyle
{\cal L} = {\cal L}_{kinetic} + {\cal L}_{axial} + {\cal L}_{ct}
&
\end{eqnarray*}
The kinetic part of lagrangian 
\begin{eqnarray*}
{{\cal L}_{kinetic}} &=& - {\rm Tr}[{\overline H}^a (i v \cdot D_{ba}-\delta_{H}\delta_{ab})H_{b}] \\
                     & & + {\rm Tr}[{\overline S}^a (i v \cdot D_{ba}-\delta_{S}\delta_{ab})S_{b}]
\end{eqnarray*}
where $\delta{H}$ and $\delta{S}$ are residual masses of H and S fields. The
axial coupling Lagrangian is: 
\begin{eqnarray*}
{\cal L}_{axial} &=& g {\rm Tr}[{\overline H}_a H_b {\slash A}_{ba} \gamma_5] \\
                 & & +g' {\rm Tr}[{\overline S_a} S_b {\slash A}_{ba} \gamma_5] \\ 
                 & & +h {\rm Tr}[{\overline H}_a S_b {\slash A}_{ba} \gamma_5 + h.c.] 
\end{eqnarray*}
where $g,\,
 g'$ are coupling constants in the ground state and in excited state 
doublets respectively,
and $h$ is the coupling between mesons belonging to different doublets.
The mass counterterm Lagrangian is:
\begin{eqnarray*}
{\cal L}_{ct} &=& Tr[(a_H {\overline H}_a H_b - a_S {\overline S_a} S_b) 
(\xi m_q \xi + \xi^{\dagger} m_q \xi^{\dagger})_{ab}] \\
              &+& Tr[(\sigma_H {\overline H_a} H_a - 
\sigma_S {\overline S_a} S_a) 
(\xi m_q \xi + \xi^{\dagger} m_q \xi^{\dagger})_{bb}]
\end{eqnarray*}
where $m_q$ = diag $(m_u, m_d, m_s)$ , $\xi^2 = \exp(2i \phi/f)$
with $\phi$ being usual matrix of pseudo-Goldstone bosons and $f \approx 130
{\rm MeV}$. In terms of heavy quark symmetry conserving and symmetry violating
terms the above lagrangian can be written as
\begin{eqnarray*}\label{finallagrang}
{\cal L}_v^{\rm ct} &=& -{\Delta_H \over 8} {\rm Tr}
[\overline H_a \sigma^{\mu \nu} H_a \sigma_{\mu \nu}] +
{\Delta_S \over 8} {\rm Tr}[\overline S_a \sigma^{\mu \nu} S_a \sigma_{\mu \nu}] \, \nonumber \\
&&+ a_H {\rm Tr} [\overline H_a H_b]\,m^\xi_{ba} - a_S {\rm Tr} [\overline S_a S
_b] \, m^\xi_{ba} \\
&& + \sigma_H {\rm Tr} [\overline H_a H_a] \, m^\xi_{bb} -
\sigma_S {\rm Tr} [\overline S_a S_a] \, m^\xi_{bb}\, \nonumber \\
&&
-{\Delta_H^{(a)} \over 8} {\rm Tr} [\overline H_a \sigma^{\mu \nu} H_b
\sigma_{\mu \nu}] \, m^{\xi}_{ba} +
{\Delta_S^{(a)} \over 8} {\rm Tr} [\overline
S_a \sigma^{\mu \nu} S_b \sigma_{\mu \nu} ] \, m^\xi_{ba}  \nonumber \\
&& - {\Delta_H^{(\sigma)} \over 8}
    {\rm Tr} [\overline H_a \sigma^{\mu \nu} H_a \sigma_{\mu \nu}]\, m^\xi_{bb}
+
{\Delta_S^{(\sigma)} \over 8}
    {\rm Tr} [\overline S_a \sigma^{\mu \nu} S_a \sigma_{\mu \nu}]\, m^\xi_{bb}
\, ,
\end{eqnarray*}
where $m^\xi_{ba} = \frac{1}{2}(\xi m_q \xi + \xi^\dagger m_q
\xi^\dagger)_{ba}$. Here $\Delta_H$ , $\Delta_S$ are symmetry(spin) violating
opertors giving rise to hyperfine splitting and $a_H$, $a_S$, $\sigma_H$,
$\sigma_S$ preserve spin-symmetry while other opertors violate heavy quark spin
symmetry.

\end{document}